%% file: main.tex
\documentclass{article}
\usepackage[T1]{fontenc} 
\usepackage[utf8]{inputenc} 
\usepackage{ismir,amsmath}
\usepackage[nosort]{cite}
\usepackage{url}
\usepackage{graphicx}
\usepackage{color}

\title{Codified audio language modeling learns useful representations for music information retrieval}

\threeauthors
  {Rodrigo Castellon\textsuperscript{*}\thanks{\textsuperscript{*} Equal contribution}} {Stanford University}
  {Chris Donahue\textsuperscript{*}} {Stanford University}
  {Percy Liang} {Stanford University}

\def\authorname{Rodrigo Castellon, Chris Donahue, Percy Liang}

\usepackage[bookmarks=false,pdfauthor={\authorname},pdfsubject={\papersubject},hidelinks]{hyperref}

\usepackage{siunitx}
\usepackage{booktabs}
\usepackage{cleveref}
\usepackage{enumitem}

\newif\ifcomment
\commenttrue
\input{macro.tex}

\begin{document}

\maketitle

\begin{abstract}
We demonstrate that language models pre-trained on codified (discretely-encoded) music audio learn representations that are useful for downstream MIR tasks. 
Specifically, we explore representations from Jukebox~\cite{dhariwal2020jukebox}: 
a music generation system containing a language model trained on codified audio from $1$M songs. 
To determine if Jukebox's representations contain useful information for MIR, 
we use them as input features to train shallow models on several MIR tasks. 
Relative to representations from conventional MIR models which are pre-trained on tagging, 
we find that using representations from Jukebox as input features yields 
$30$\% 
stronger performance on average across four MIR tasks: 
tagging, genre classification, key detection, and emotion recognition.
For key detection, we observe that representations from Jukebox are considerably stronger than those from models pre-trained on tagging, 
suggesting that pre-training via codified audio language modeling may address blind spots in conventional approaches. 
We interpret the strength of Jukebox's representations as evidence that modeling audio instead of tags provides richer representations for MIR.
\end{abstract}

\section{Introduction}\label{sec:introduction}

It is conventional in MIR\footnote{MIR has a broad definition, but in this paper ``MIR'' refers specifically to making discriminative predictions on music audio.} 
to \emph{pre-train} models on large labeled datasets for one or more tasks (commonly tagging), and reuse the learned representations for different \emph{downstream} tasks~\cite{hamel2013transfer,oord2014transfer,choi2017transfer,park2017representation,lee2018samplecnn,lee2019representation,pons2019musicnn,huang2020large,kim2020one}. 
Such \emph{transfer learning} approaches decrease the amount of labeled data needed to perform well on downstream tasks, which is particularly useful in MIR where labeled data for many important tasks is scarce~\cite{mcfee2018open,chen2019data}. 
Historically-speaking, 
improvement on downstream tasks is enabled by finding ever-larger sources of labels for pre-training---in chronological order: 
tags~\cite{oord2014transfer},  
metadata~\cite{park2017representation,lee2019representation,huang2020large,kim2020one}, and 
recently, co-listening data~\cite{huang2020large}.
However, it stands to reason that directly modeling music~\emph{audio} (as opposed to labels) could yield richer representations. 
Recently, contrastive learning~\cite{chen2020simple} has been proposed as an MIR pre-training strategy which learns representations from audio~\cite{spijkervet2021contrastive}, 
but this paradigm has yet to exceed the performance of label-based pre-trained models on downstream tasks.

Outside of the discriminative MIR landscape, 
a recent system called Jukebox~\cite{dhariwal2020jukebox} demonstrated promising performance for generating music audio. 
To achieve this result, Jukebox leverages recent architectural developments from natural language processing (NLP) by \emph{codifying} audio---encoding high-rate continuous audio waveforms into lower-rate discrete sequences which can be fed in directly to NLP models. 
Specifically, Jukebox trains a Transformer~\cite{vaswani2017attention,child2019generating} \emph{language model}, 
an autoregressive generative model, 
on codified audio from $1$M songs. 
Purely for convenience, we refer to Jukebox's training procedure as \emph{codified audio language modeling} (\calm). 

\begin{figure*}
    \centering
    \includegraphics[width=0.99\linewidth]{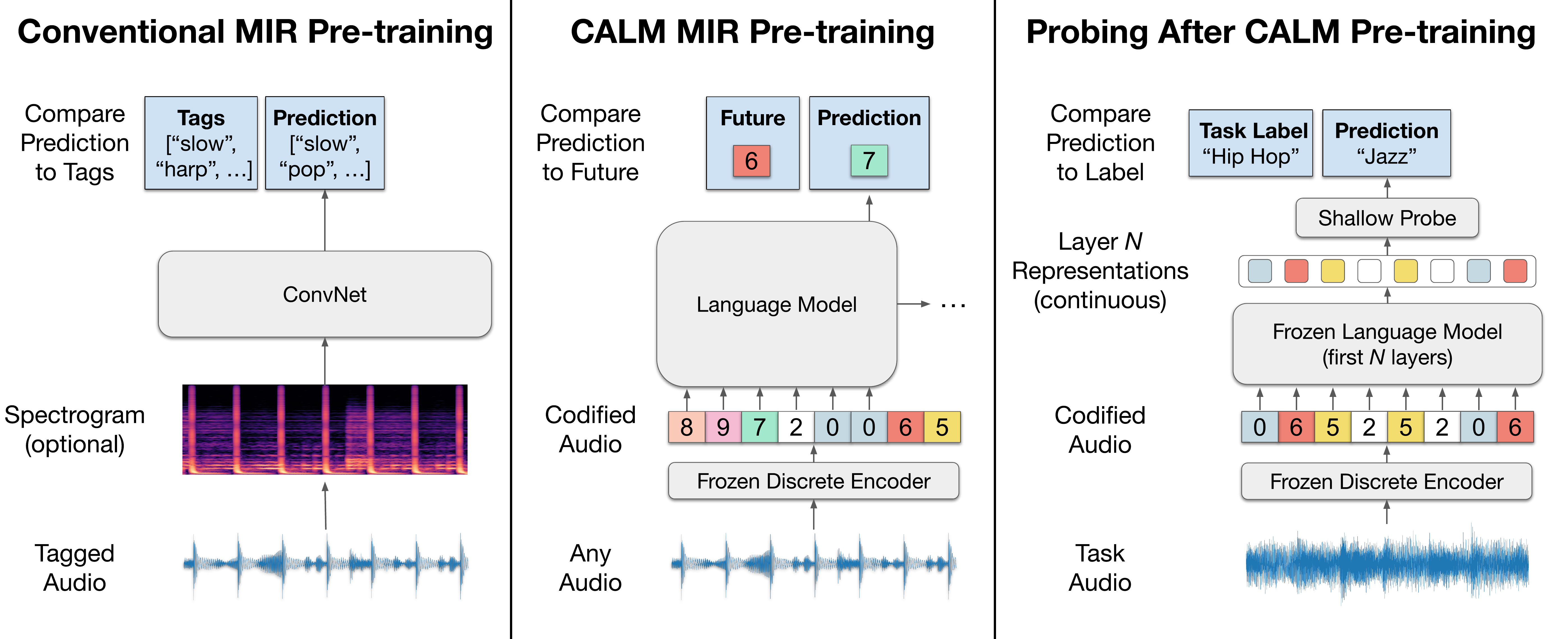}
    \caption{
    Conventional MIR pre-training (\textbf{left}) trains convolutional neural networks on audio spectrograms using manually-annotated labels from tagging datasets. 
    In contrast, \calm{} MIR pre-training (\textbf{middle}) involves training a language model on codified audio, 
    which has been previously explored for music generation~\cite{dieleman2018challenge,dhariwal2020jukebox}---here, we propose to use it for discriminative MIR tasks. 
    To determine if \calm{} pre-training is effective for MIR, we probe for information about particular MIR tasks (\textbf{right}) in resultant representations. 
    Specifically, we extract features from the learned language model for the audio in small, task-specific labeled datasets, and use these features to train shallow probing models on each task.}
    \label{fig:one}
\end{figure*}

While Jukebox already demonstrates that \calm{} is useful for music \emph{generation}, 
in this work we demonstrate that \calm{} is also useful as a pre-training procedure for \emph{discriminative} MIR tasks. 
To this end, 
we repurpose Jukebox for MIR by first using it to extract audio feature representations, 
and then training shallow models (\emph{probes}~\cite{alain2016understanding,hupkes2018visualisation}) on downstream tasks using these features as input (\Cref{fig:one}). 
Relative to representations from models pre-trained with tagging, 
we find that representations from Jukebox are $30$\% more effective on average 
when used to train probes on four downstream MIR tasks: tagging, genre classification, key detection, and emotion recognition. 
We also observe that representations from Jukebox are much more useful for key detection than those from models pre-trained on tagging, 
which suggests that \calm{} pre-training may be particularly beneficial for tasks which have little to do with tagging. 
This simple setup of training shallow models on representations from Jukebox is even competitive with purpose-built state-of-the-art methods on several tasks.

To facilitate reproducibility and encourage further investigation of these representations and tasks~\cite{mcfee2018open}, 
we release all of our code for this project, 
alongside images for Docker containers which provide full provenance for our experiments.\footnote{
Code: \url{https://github.com/p-lambda/jukemir} \\
Containers: \url{https://hub.docker.com/orgs/jukemir} \\
All experiments reproducible on the CodaLab platform: \\
\url{https://worksheets.codalab.org/worksheets/0x7c5afa6f88bd4ff29fec75035332a583}
}
We note that, 
while \calm{} pre-training at the scale of Jukebox requires substantial computational resources, 
our post hoc experiments with Jukebox 
only require a single commodity GPU with $12$ GB memory.

\section{\calm{} Pre-training}

\calm{} was first proposed by van~den~Oord~\etal{} and used for unconditional speech generation~\cite{oord2017neural}. 
As input, \calm{} takes a collection of raw audio waveforms (and optionally, conditioning metadata), 
and learns a distribution $p(\text{audio} \mid \text{metadata})$. 
To this end, \calm{} adopts a three-stage approach: 
(1)~\emph{codify} a high-rate continuous audio signal into lower-rate discrete codes, 
(2)~train a \emph{language model} on the resulting codified audio and optional metadata, i.e., learn $p(\text{codified audio} \mid \text{metadata})$, 
and 
(3)~decode sequences generated by the language model to raw audio.\footnote{This third stage is not necessary for transfer learning.} 
The original paper~\cite{oord2017neural} also proposed a strategy for codifying audio called the vector-quantized variational auto-encoder (VQ-VAE), 
and the language model was a WaveNet~\cite{oord2016wavenet}. 
Within music, 
\calm{} was first used by Dieleman~\etal{} for unconditional piano music generation~\cite{dieleman2018challenge}, 
and subsequently, 
Dhariwal~\etal{} used \calm{} to build a music generation system called Jukebox~\cite{dhariwal2020jukebox} with conditioning on genre, artist, and optionally, lyrics. 

Despite promising results on music audio generation, 
\calm{} has not yet been explored as a pre-training strategy for discriminative MIR.
We suspect that effective music audio generation
necessitates 
intermediate representations that would also contain useful information for MIR. 
This hypothesis is further motivated by an abundance of previous work in NLP suggesting that generative and self-supervised pre-training can yield powerful representations for discriminative tasks~\cite{peters2018deep,devlin2018bert,radford2019language,liu2021gpt}.

To explore this potential,
we repurpose Jukebox for MIR. 
While Jukebox was designed only for generation, 
its internal language model was trained on codified audio from a corpus of $1.2$M songs from many genres and artists, 
making its representations potentially suitable for a multitude of downstream MIR tasks. 
Jukebox consists of two components---the first is 
a small~($2$M parameters) VQ-VAE model~\cite{oord2017neural} that learns to codify high-rate~(\SI{44.1}{\kilo\hertz}), continuous audio waveforms into lower-rate~($\sim$\SI{345}{\hertz}), discrete code sequences with a vocabulary size of $2048$ ($11$~bits). 
The second component is 
a large~($5$B parameters) language model that learns to generate codified audio using a Transformer decoder---an architecture originally designed for modeling natural language~\cite{vaswani2017attention,child2019generating}. 
By training on codified audio (as in~\cite{dieleman2018challenge,dhariwal2020jukebox}) instead of raw audio (as in~\cite{oord2016wavenet,child2019generating}), 
language models are (empirically) able to learn longer-term structure in music, 
while simultaneously using significantly less memory to model the same amount of audio. 

Like conventional MIR models which pre-train on tagging and/or metadata, 
Jukebox also makes use of genre and artist labels during training, 
providing them as conditioning information to allow for increased user control over the music generation process. 
Hence, while CALM in general is an unsupervised strategy that does not require labels, 
transfer learning from Jukebox specifically should \emph{not} be considered an unsupervised approach (especially for downstream tasks like genre detection). 
However, 
by modeling the \emph{audio} itself instead of modeling the \emph{labels} (as in conventional MIR pre-training), we hypothesize that Jukebox learns richer representations for MIR tasks than conventional strategies.

\section{Extracting suitable representations from Jukebox}

Here we describe how we extract audio representations from Jukebox which are suitable as input features for training shallow models. 
While several pre-trained Jukebox models exist with different sizes and conditioning information, 
here we use the $5$B-parameter model without lyrics conditioning (named ``5b''), 
which is a sparse transformer~\cite{vaswani2017attention,child2019generating} containing $72$ layers. 
Each layer yields $4800$-dimensional activations for each element in the codified audio sequence, i.e., approximately $345$ times per second. 
To extract representations from this model for a particular audio waveform, we 
(1)~resample the waveform to $44.1$kHz, 
(2)~normalize it, 
(3)~codify it using the Jukebox VQ-VAE model, and 
(4)~input the codified audio into the language model, interpreting its layer-wise activations as representations. 
Jukebox was trained on ${\sim}24$-second audio clips (codified audio sequences of length $8192$)---we feed in this same amount of audio at a time when extracting representations. 
In addition to the genre and artist conditioning fields mentioned previously, 
Jukebox expects two additional fields: total song length and clip offset---to ensure that representations only depend on the input audio, 
we simply pass in ``unknown'' for artist and genre, one minute for song length, and zero seconds for clip offset.\footnote{We observed in initial experiments that passing in ground-truth conditioning information had little effect on downstream performance. Hence, we elected to pass in placeholder metadata to maintain the typical type signature for audio feature extraction (audio as the only input).}

\begin{figure}
    \centering
    \includegraphics[width=0.95\linewidth]{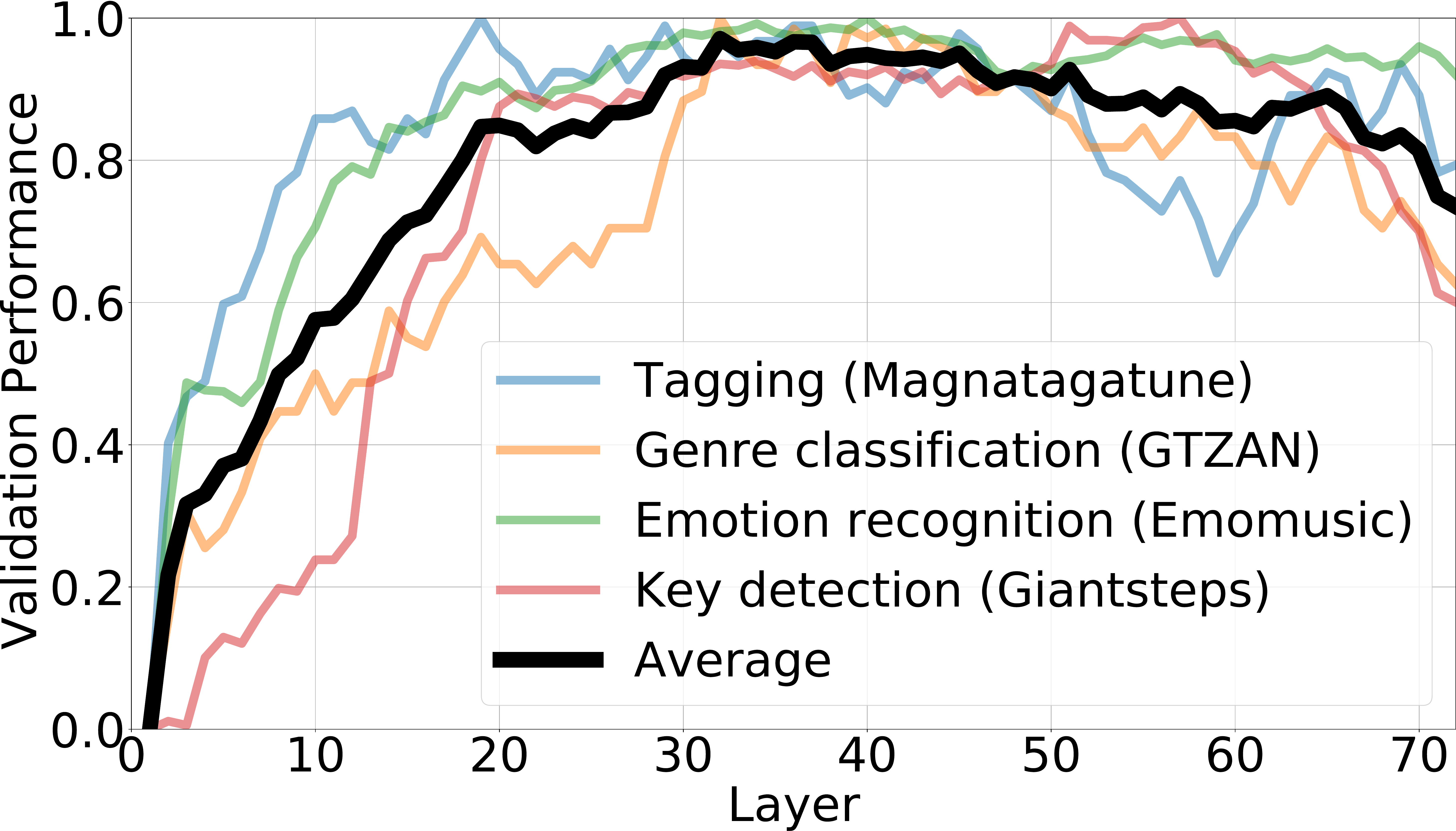}
    \caption{Normalized validation performance of linear models trained on representations from specific layers of Jukebox across four downstream MIR tasks. On average, the strongest representations for these tasks come from the middle of Jukebox.}
    \label{fig:layers}
\end{figure}

The Jukebox language model yields an unwieldy amount of data---for every $24$-second audio clip, it emits $24\times345\times72\times4800$ numbers, 
i.e., 
over $10$GB if stored naively as $32$-bit floating point. 
We reduce the amount of data by mean pooling across time, 
a common strategy in MIR transfer learning~\cite{choi2017transfer,pons2019musicnn}, 
which aggregates more than $10$GB of activations to around $1$MB ($72\times4800$).

\subsection{Layer selection}\label{sec:layersel}

While pooling across time dramatically reduced the dimensionality of Jukebox's outputs, 
training shallow classifiers on $72\times4800$ features is still computationally expensive. 
To further reduce the dimensionality, 
we use only one of the layers from Jukebox---the middle layer ($36$)---yielding a total of $4800$ features per $24$ second audio clip. 
Unlike conventional pre-training, 
where the strongest representations for transfer learning typically lie at the end of the model~\cite{zeiler2014visualizing}, 
the strongest representations from pre-trained language models tend to lie towards the middle of the network~\cite{liu2019linguistic,chen2020generative,chi2020finding,rogers2020primer}. 
To confirm this observation in our context, 
we trained linear models using representations from different layers of Jukebox on our downstream MIR tasks---average performance indeed peaked at the middle layers~(\Cref{fig:layers}). 

In addition to using the middle layer, 
we experimented with two other layer selection strategies: 
(1)~sub-sampling layers across the network, 
and 
(2)~selecting relevant layers in a task-specific fashion.\footnote{This procedure selected layers that were the most jointly informative in a greedy fashion, measured by task performance with a linear probe.} 
We found that the simplest strategy of using only the middle layer was equally effective and more computationally practical\footnote{While the entirety of Jukebox does \emph{not} fit on a single commodity GPU with $12$GB memory, the first $36$ layers \emph{do} fit.} than the other two layer selection strategies.

\section{Downstream task descriptions}\label{sec:tasks}

\begin{table}[t]
\centering
\begin{tabular}{lccc}
\toprule
Task & Size & Metrics & \#Out \\
\midrule
Tagging~\cite{law2009evaluation} & $25860$ & AUC/AP & 50 \\
Genre classification~\cite{kereliuk2015deep} & $930$ & Accuracy & 10 \\
Key detection~\cite{knees2015two} & $1763$ & Score & 24 \\
Emotion recognition~\cite{soleymani20131000} & $744$ & A/V $R^2$ & 2 \\
\bottomrule
\end{tabular}
\caption{Basic information about the four tasks we consider in this work, including the size of each task-specific dataset in terms of number of labeled examples, relevant metrics for each task, and the number of model outputs required for each dataset.}
\label{tab:tasks}
\end{table}

We select four downstream MIR tasks to constitute a benchmark for comparing different audio feature representations: 
(1)~tagging,
(2)~genre classification,
(3)~key detection, and
(4)~emotion recognition.
A summary of the datasets used for each task appears in~\Cref{tab:tasks}. 
These tasks were selected to cover a wide range of dataset sizes ($744$ examples for emotion recognition vs. $26$k examples for tagging) 
and subjectivity (emotion recognition is more subjective vs. key detection is more objective). 
Additionally, each task has an easily-accessible dataset with standard evaluation criteria. 
We describe each of these tasks and metrics below.

\subsection{Tagging}

Tagging involves determining which tags from a fixed set of tags apply to a particular song. 
Categories of tags include 
genre~(e.g.,~jazz), 
instrumentation~(e.g.,~violin), 
emotions~(e.g.,~happy), and 
characteristics~(e.g.,~fast). 
There are two large datasets for tagging, which both contain human-annotated tags for $30$-second clips:  
MagnaTagATune~\cite{law2009evaluation}~(MTT) which contains around $26$k clips, and 
a tagged subset of $240$k clips from the Million Song Dataset~\cite{bertin2011million}~(MSD). 
While both datasets contain a large vocabulary of tags, 
typical usage involves limiting the vocabulary to the $50$ most common tags in each.

Because it is the largest non-proprietary MIR dataset, MSD is commonly used for pre-training models for transfer learning. 
To mitigate an unfair advantage of methods which pre-train on MSD, 
we use MTT instead of MSD to benchmark representations on tagging performance. 
While both datasets are superficially similar (choosing from $50$ tags for $30$-second clips), 
their label distributions are quite different: MSD is skewed towards genre tags, while MTT is skewed towards instrumentation tags.

We use the standard ($12$:$1$:$3$) train, validation, and test split for MTT~\cite{oord2014transfer}. 
Additionally, we report both common metrics (both are macro-averaged over tags as is conventional): area under the receiver operating characteristic curve (\mttauc), and average precision (\mttap).\footnote{Most past work refers to the quantity of average precision as area under the precision-recall curve.}
We note that inconsistencies in handling unlabeled examples for past work on MTT have been observed~\cite{won2020evaluation}---some work discards examples without top-$50$ tags during training, evaluation, or both. 
In this work, we do not discard any examples.

\subsection{Genre classification}

Genre classification involves assigning the most appropriate genre from a fixed list for a given song. 
For this task, we report accuracy on the GTZAN dataset~\cite{tzanetakis2002musical}, which contains $30$-second clips from $10$ distinct genres. 
We adopt the ``fault-filtered'' split from \cite{kereliuk2015deep} which addresses some of the reported issues with this dataset~\cite{sturm2013gtzan}. 
We note that this task has a high degree of overlap with tagging, as tagging datasets typically have a number of genres within their tag vocabulary. 
In fact, seven of ten genres in GTZAN are present in the tag list of MSD.

\subsection{Key detection}

Key detection involves predicting both the scale and tonic pitch class for the underlying key of a song. 
We investigate the Giantsteps-MTG and Giantsteps datasets~\cite{knees2015two} which include songs in major and minor scales for all pitch classes, i.e., a $24$-way classification task. 
As in past work~\cite{korzeniowski2017end}, we use the former for training and the latter for testing. 
Because no standard validation split exists for Giantsteps-MTG, 
we follow~\cite{kereliuk2015deep} and create an artist-stratified $4$:$1$ split for training and validation, 
which we include in our codebase for reproducibility. 
The music in this dataset is all electronic dance music, 
and the clips are two minutes in length. 
We report the typical weighted score metric for Giantsteps (\gssco): an accuracy measure which gives partial credit for reasonable mistakes such as predicting the relative minor key for the major ground truth~\cite{raffel2014mireval}. 

\subsection{Emotion recognition}\label{sec:emomusic}

Emotion recognition involves predicting human emotional response to a song. 
Data is collected by asking humans to report their emotional response on a two dimensional valence-arousal plane~\cite{huq2010automated}, 
where valence indicates positive versus negative emotional response, and arousal indicates emotional intensity. 
We use the Emomusic dataset~\cite{soleymani20131000}, 
which contains $744$ clips of $45$ seconds in length. 
We investigate the static version of this task where original time-varying annotations are averaged together to constitute a clip-level annotation. 
Because this dataset does not have a standard split, it is difficult to directly compare with past work. 
To simplify comparison going forward, we created an artist-stratified split of Emomusic, which is released in our codebase. 
We take the highest reported numbers from past work to characterize ``state-of-the-art'' performance, though we note that these numbers are not directly comparable to our own due to differing splits. 
We report the coefficient of determination between the model predictions and human annotations for arousal~(\emoa) and valence~(\emov).

\section{Probing experiments}

Here we describe our protocol for probing for information about MIR tasks in representations from Jukebox and other pre-trained models, 
i.e., 
measuring performance of shallow models trained on these tasks using different representations as input features. 
We borrow the term ``probing'' from analogous investigations in NLP~\cite{hupkes2018visualisation,conneau2018you,hewitt2019structural}, 
however such methodology is common in transfer learning for MIR~\cite{hamel2013transfer,oord2014transfer,choi2017transfer,park2017representation,lee2019representation,pons2019musicnn,huang2020large,kim2020one}. 

\subsection{Descriptions of representations}

\newcommand{\tabreprow}[4]{ #2 & #1 & #4 \\}

\begin{table}[t]
\centering
\begin{tabular}{llc}
\toprule
\tabreprow{Pre-training strategy}{Representation}{Rate}{Dimensions}
\midrule
\tabreprow{N/A}{\repchroma}{}{$72$}
\tabreprow{N/A}{\repmfcc}{\SI{23.4}{\hertz}}{$120$}
\tabreprow{MSD Tagging~\cite{oord2014transfer}}{\repchoi~\cite{choi2017transfer}}{\SI{0.0345}{\hertz}}{$160$}
\tabreprow{MSD Tagging~\cite{oord2014transfer}}{\repmusicnn~\cite{pons2019musicnn}}{\SI{62.5}{\hertz}}{$4194$}
\tabreprow{Contrastive~\cite{chen2020simple}}{\repclmr~\cite{spijkervet2021contrastive}}{\SI{0.373}{\hertz}}{$512$}
\tabreprow{\calm{}~\cite{oord2017neural}}{\repjuke~\cite{dhariwal2020jukebox}}{\SI{345}{\hertz}}{$4800$}
\bottomrule
\end{tabular}
\caption{Basic statistics about the six representations we examine in this work.}
\label{tab:representations}
\end{table}

\newcommand{\tabresrow}[8]{ #1 & #3 & #4 & #2 & #7 & #5 & #6 & #8 \\}

\begin{table*}[t]
\centering
\begin{tabular}{lccccccc}


\toprule

& \multicolumn{2}{c}{Tags} & \multicolumn{1}{c}{Genre} & \multicolumn{1}{c}{Key} & \multicolumn{2}{c}{Emotion} \\
\cmidrule(lr){2-3} \cmidrule(lr){4-4} \cmidrule(lr){5-5} \cmidrule(lr){6-7}

\tabresrow{Approach}{\gtzanacc}{\mttauc}{\mttap}{\emoa}{\emov}{\gssco}{\gmueavg}

\midrule

\tabresrow{(No pre-training) Probing \repchroma{}}                                  {$32.8$}{$77.6$}{$18.5$}{$29.3$}{ $5.9$}{$56.5$}{$38.7$}
\tabresrow{(No pre-training) Probing \repmfcc{}}                                    {$44.8$}{$85.8$}{$30.2$}{$47.9$}{$26.5$}{$14.6$}{$38.7$}
\tabresrow{(Tagging) Probing \repchoi{}~\cite{choi2017transfer}}                    {$75.9$}{$89.7$}{$36.4$}{$67.3$}{$43.4$}{$13.1$}{$51.9$}
\tabresrow{(Tagging) Probing \repmusicnn{}~\cite{pons2019musicnn}}                  {$79.0$}{$90.6$}{$38.3$}{$70.3$}{$46.6$}{$12.8$}{$53.7$}
\tabresrow{(Contrastive) Probing \repclmr{}~\cite{spijkervet2021contrastive}}       {$68.6$}{$89.4$}{$36.1$}{$67.8$}{$45.8$}{$14.9$}{$50.8$}
\tabresrow{(\calm) Probing \repjuke~\cite{dhariwal2020jukebox}}                     {$\mathbf{79.7}$}{$\mathbf{91.5}$}{$\mathbf{41.4}$}{$\mathbf{72.1}$}{$\mathbf{61.7}$}{$\mathbf{66.7}$}{$\mathbf{69.9}$}

\midrule
\tabresrow{State-of-the-art~\cite{huang2020large,pons2019musicnn,lee2018samplecnn,weninger2014line,koh2021comparison,pioneer2015rekordbox}}{$\mathbf{82.1}$}{$\mathbf{92.0}$}{$38.4$}{$70.4$*}{$55.6$*}{$\mathbf{79.6}$}{$\mathbf{72.5}$*}
\tabresrow{~~~~Pre-trained~\cite{huang2020large,spijkervet2021contrastive,lee2018samplecnn,koh2021comparison,koh2021comparison,jiang2019mirex}}{$\mathbf{82.1}$}{$\mathbf{92.0}$}{$35.9$}{$67.1$*}{$55.6$*}{$75.8$}{$70.8$*}
\tabresrow{~~~~From scratch~\cite{pons2019musicnn,pons2019musicnn,medhat2017masked,weninger2014line,weninger2014line,korzeniowski2017end}}{$65.8$}{$90.7$}{$38.4$}{$70.4$*}{$50.0$*}{$74.3$}{$66.2$*}

\bottomrule

\end{tabular}
\caption{Comparing performance of probes on representations from a model pre-trained with \calm{} to other pre-trained MIR models (top section) to reported state-of-the-art performance (bottom section) across four tasks: 
(1)~tagging (\mttauc/\mttap), 
(2)~genre classification (\gtzanacc), 
(3)~key detection (\gssco), 
and 
(4)~emotion recognition (\emoa/\emov).
For all six metrics, the max score is $100$ and higher is better---see \Cref{sec:tasks} for a full description of tasks/metrics. 
For each metric, the best probing-based approach and the best approach overall are \textbf{bolded}. 
We also report an average score across all four tasks; tasks with multiple evaluation metrics are averaged beforehand. 
On all metrics, probing \repjuke{} is more effective than probing representations from other pre-trained models. 
Probing \repjuke{} is competitive with task-specific state-of-the-art approaches for all tasks/metrics except key detection (\gssco). 
Note that the ordering of citations in the bottom section corresponds to respective column ordering.
* indicates that past work on Emomusic evaluates on different subsets of the dataset than our work and hence numbers are not directly comparable---see~\Cref{sec:emomusic} for details.
}
\label{tab:results}
\end{table*}

In addition to probing representations from Jukebox (an exemplar of \calm{} pre-training), 
we probe four additional representations which are emblematic of three other MIR pre-training strategies (\Cref{tab:representations}). 
Before pre-training, hand-crafted features were commonplace in MIR---as archetypal examples, 
we probe 
constant-Q chromagrams (\repchroma)
and
Mel-frequency cepstral coefficients (\repmfcc), extracted with librosa~\cite{mcfee2015librosa} using the default settings. 
As in~\cite{choi2017transfer}, we concatenate the mean and standard deviation across time of both the features and their first- and second-order discrete differences.  
We also probe two examples of the current conventional paradigm which pre-trains on tagging using MSD: a convolutional model proposed by Choi~\etal{}~\cite{choi2017transfer} (\repchoi), and a more modern convolutional model from~\cite{pons2019musicnn} (\repmusicnn). 
Finally, we compare to a recently-proposed strategy for MIR pre-training called \emph{contrastive learning of musical representations}~\cite{spijkervet2021contrastive} (\repclmr), though we note that the only available pre-trained model from this work was trained on far less audio (a few thousand songs) than the other pre-trained models (\repchoi{}, \repmusicnn{}, and \repjuke{}). 

All of these strategies operate at different frame rates,~i.e.,~they produce a different number of representation vectors for a fixed amount of input audio. 
To handle this, we follow common practice of mean pooling representations across time~\cite{choi2017transfer,pons2019musicnn}. 
While \repchroma{}, \repmfcc{}, and \repclmr{} produce a single canonical representation per frame, 
we note that the other three produce multiple representations per frame, i.e., the outputs of individual layers in each model. 
For \repchoi, we concatenate all layer representations together, which was shown to have strong performance on all downstream tasks in~\cite{choi2017transfer}. 
For \repmusicnn, we concatenate together the mean and max pool of three-second windows (before mean pooling across these windows), i.e., the default configuration for that approach.
For \repjuke, 
we use the middle layer of the network as motivated in~\Cref{sec:layersel}. 
By using a single layer, we also mitigate the potential of a superficial dimensionality advantage for \repjuke{}, as this induces a dimensionality similar to that of \repmusicnn{} ($4800$ and $4194$ respectively; see \Cref{tab:representations}).

Unlike other representations which operate on short context windows, 
\repchoi{} and \repjuke{} were trained on long windows of $29$ seconds and $24$ seconds of audio respectively. 
Accordingly, 
for the three datasets with short clips (tagging, genre classification, and emotion recognition all have clips between $30$ and $45$ seconds in length), 
we adopt the policy from~\cite{choi2017transfer} and simply truncate the clips to the first window when computing representations for \repchoi{} and \repjuke{}. 
Because clips from the key detection dataset are much longer (two minutes), 
we split the clips into $30$-second windows for all methods and train probes on these shorter windows. 
At test time, we ensemble window-level predictions into clip-level predictions before computing the score. 

\subsection{Probing protocol}

To probe representations for relevant information about downstream MIR tasks, 
we train shallow supervised models (linear models and one-layer MLPs) on each task using these representations as input features. 
As some representations may require different hyperparameter configurations for successful training, 
we run a grid search over the following hyperparameters ($216$ total configurations) 
for each representation and task ($24$ total grid searches), 
using early stopping based on task-specific metrics computed on the validation set of each task: 
\begin{itemize}[nolistsep]
    \item Feature standardization: \{off, on\}
    \item Model: \{Linear, one-layer MLP with $512$ hidden units\}
    \item Batch size: \{$64$, $256$\}
    \item Learning rate: \{$1\mathrm{e}$-$5$, $1\mathrm{e}$-$4$, $1\mathrm{e}$-$3$\}
    \item Dropout probability: \{$0.25$, $0.5$, $0.75$\}
    \item L$2$ regularization: \{$0$, $1\mathrm{e}$-$4$, $1\mathrm{e}$-$3$\}
\end{itemize}

While we use this same hyperparameter grid for all tasks, 
the learning objective varies by task 
(cross-entropy for genre classification and key detection, 
independent binary cross-entropy per tag for tagging, and 
mean squared error for emotion recognition) 
as does the number of probe outputs (\Cref{tab:tasks}). 
Some tasks have multiple metrics---we early stop on \mttauc{} for tagging as it is a more common metric than \mttap{}, and on the average of \emoa{} and \emov{} for emotion recognition. 
We take the model with the best early stopping performance from each grid search and compute its performance on the task-specific test set.

\section{Results and Discussion}

In~\Cref{tab:results}, we report performance of all representations on all tasks and metrics, as well as average performance across all tasks. 
Results are indicative that \calm{} is a promising paradigm for MIR pre-training. 
Specifically, we observe that probing the representations from \repjuke{} (learned through \calm{} pre-training) achieves an average of $69.9$, 
which is $30$\% higher relative to the average of the best representation pre-trained with tagging (\repmusicnn{} achieves an average of $53.7$). 
Performance of \repjuke{} on all individual metrics is also higher than that of any other representation. 
Additionally, \repjuke{} achieves an average performance that is $38$\% higher than that of \repclmr{}. 
Representations from all pre-trained models outperform hand-crafted features (\repchroma{} and \repmfcc) on average.
Note that these results are holistic comparisons across different model architectures, model sizes, and amounts of pre-training data (e.g.,~\repclmr{} was trained on far less data than \repjuke{}), and hence not sufficient evidence to claim that \calm{} is the ``best'' music pre-training strategy in general.

We also observe that \repjuke{} contains substantially more information relevant for key detection than other representations. 
While \repchroma{} (spectrogram projected onto musical pitch classes) contains information relevant to key detection by design, 
all other representations besides \repjuke{} yield performance on par with that of a majority classifier (outputting ``F minor'' for every example scores $15.0$)---hence, these representations contain almost no information about this task. 
For models pre-trained with tagging (\repchoi{} and \repmusicnn{}), 
intuition suggests that this is because 
none of the tags in MSD relate to key signature. 
For \repclmr{}, 
we speculate that the use of transposition as a data augmentation strategy 
also results in a model that contains little useful information about key signature. 
While tagging and CLMR were not designed with the intention of supporting transfer to key detection, we argue that it is generally desirable to have a unified music representation which performs well on a multitude of downstream MIR tasks.
Hence, we interpret the comparatively stronger performance of \repjuke{} on key detection as evidence that \calm{} pre-training addresses blind spots present in other MIR pre-training paradigms.

In the bottom section of~\Cref{tab:results}, we also report state-of-the-art performance for purpose-built methods on all tasks, which is further broken down by models which use any form of pre-training (including pre-training on additional task-specific data as in~\cite{jiang2019mirex}) vs. ones that are trained from scratch. 
Surprisingly, 
we observe that probing \repjuke{} is competitive with state-of-the-art for all tasks except for key detection, 
and achieves an average only $4$\% lower relative to that of state-of-the-art. 
On tagging, probing \repjuke{} achieves similar \mttauc{} to a strategy which pre-trains on a proprietary dataset of $10$M songs using supervision~\cite{huang2020large}. 
We interpret the strong performance of this simple probing setup as evidence that \calm{} pre-training is a promising path towards models that are useful for many MIR tasks. 

We believe that \calm{} pre-training is promising for MIR not just because of the strong performance of an existing pre-trained model (Jukebox), 
but also because there are numerous avenues which may yield further improvements for those 
with the data and computational resources to explore them.
Firstly, 
\calm{} could be scaled up to pre-train even larger models on more data (Jukebox was trained on $1$M songs, while Spotify has an estimated $70$M songs in its catalog). 
In~\cite{kaplan2020scaling}, it is observed that increasing model and dataset size yields predictable improvements to cross-entropy for language modeling in NLP, 
an insight which may also hold for \calm{} pre-training for MIR. 
Secondly,
we anticipate that fine-tuning a model pre-trained with \calm{} would outperform our probing setup. 
Finally, 
taking a cue from related findings in NLP, 
we speculate that \calm{} pre-training with a bidirectional model and masked language modeling (as in BERT~\cite{devlin2018bert}) would outperform the generative setup of Jukebox (that of GPT~\cite{radford2018improving}).

\section{Related Work}

Transfer learning 
has been an active area of study in MIR for over a decade. 
An early effort seeking to replace hand-crafted features used neural networks to automatically extract context-independent features from unlabeled audio~\cite{hamel2010learning} and used those features for a supervised learning task. 
Other early efforts focused on learning shared embedding spaces between audio and metadata~\cite{weston2011multi,hamel2013transfer} or directly using outputs from pre-trained tagging models for music similarity judgements~\cite{seyerlehner2012improved}. 

The predominant strategy for MIR pre-training using large tagging datasets was first proposed by van~den~Oord~\etal~2014~\cite{oord2014transfer}. 
This work pre-trained deep neural networks on MSD and demonstrated promising performance on other tagging and genre classification tasks. 
Choi~\etal~2017~\cite{choi2017transfer} pre-trained on MSD but using a convolutional neural network and also explored a more diverse array of downstream tasks---we use their pre-trained model as one of our baselines. 
More recent improvements use the same approach with different architectures~\cite{lee2018samplecnn,pons2019musicnn}, the latest of which is another one of our baselines. 

Other strategies for MIR transfer learning have been proposed. 
Some work pre-trains on music metadata (e.g., artist, album) instead of tags~\cite{park2017representation,lee2019representation}. 
In contrast to the manual annotations required for tagging-based pre-training, metadata is much cheaper to obtain, but performance of pre-training on metadata is comparable to that of pre-training on tagging. 
Kim~\etal~2020~\cite{kim2020one} improve over Choi~\etal~2017~\cite{choi2017transfer} using a multi-task approach that pre-trains on both tags and metadata. 
Huang~\etal~\cite{huang2020large} demonstrate that metadata can be combined with proprietary co-listening data for pre-training on $10$M songs to achieve state-of-the-art performance on MTT---probing representations from \calm{} pre-training on $1$M songs achieves comparable performance on MTT (\Cref{tab:results}).  
Finally, contrastive learning~\cite{chen2020simple} has been proposed as a strategy for MIR pre-training~\cite{favory2020learning,ferraro2021enriched,spijkervet2021contrastive}---we compare to such a model from Spijkervet~and~Burgoyne~2021~\cite{spijkervet2021contrastive}.

While \calm{} has not previously been explored for MIR transfer learning, it has been explored for other purposes. 
van~den~Oord~\etal~2017~\cite{oord2017neural} first proposed \calm{} and used it for unconditional speech generation.
Variations of \calm{} 
have been used as pre-training for speech recognition~\cite{baevski2019effectiveness,baevski2020wav2vec} and urban sound classification~\cite{verma2020framework}. 
\calm{} has also been explored for music generation~\cite{dieleman2018challenge,dhariwal2020jukebox}. 
\calm{} is related to past work on language modeling of raw (i.e.,~not codified) waveforms~\cite{oord2016wavenet,mehri2016samplernn,kalchbrenner2018efficient}, 
which tends to be less effective for capturing long-term dependencies compared to modeling codified audio. 
Language models have also been used extensively for modeling symbolic music~\cite{eck2002finding,simon2017performance,huang2018music}, 
including some work on pre-training on large corpora of scores for transfer learning~\cite{donahue2019lakhnes,hung2019improving}.

\section{Conclusion}

In this work we demonstrated that \calm{} is a promising pre-training strategy for MIR. 
Compared to conventional approaches, \calm{} learns richer representations by modeling audio instead of labels. 
Moreover, \calm{} allows MIR researchers to repurpose NLP methodology---historically, repurposing methodology from another field (computer vision) has provided considerable leverage for MIR. 
Finally, \calm{} suggests a direction for MIR research where enormous models pre-trained on large music catalogs break new ground on MIR tasks, 
analogous to ongoing paradigm shifts in other areas of machine learning.

\section{Acknowledgements}

We would like to thank
Nelson~Liu, 
Mina~Lee, 
John~Hewitt, 
Janne~Spijkervet, 
Minz~Won, 
Jordi~Pons, 
Ethan~Chi, 
Michael~Xie, 
Ananya~Kumar, 
and 
Glen~Husman 
for helpful conversations about this work.
We also thank all reviewers for their helpful feedback.

\bibliography{main}

\end{document}

%% file: macro.tex
\ifcomment
\newcommand{\todo}[1]{{\color{blue} TODO: {#1}}}
\newcommand{\rc}[1]{{\color{magenta}{[RC: #1]}}}
\newcommand{\cd}[1]{{\color{blue}{[CD: #1]}}}
\newcommand{\pl}[1]{\textcolor{red}{[PL: #1]}}
\else
\newcommand{\todo}[1]{}
\newcommand{\rc}[1]{}
\newcommand{\cd}[1]{}
\newcommand{\pl}[1]{}
\fi

\newcommand{\gtzanacc}{GTZAN}
\newcommand{\mttauc}{$\text{MTT}_{\text{AUC}}$}
\newcommand{\mttap}{$\text{MTT}_{\text{AP}}$}
\newcommand{\emoa}{$\text{Emo}_{\text{A}}$}
\newcommand{\emov}{$\text{Emo}_{\text{V}}$}
\newcommand{\gssco}{GS}
\newcommand{\gmueavg}{Average}

\newcommand{\calm}{CALM}

\newcommand{\rep}[1]{\textsc{#1}}
\newcommand{\repmfcc}{\rep{MFCC}}
\newcommand{\repchroma}{\rep{Chroma}}
\newcommand{\repclmr}{\rep{CLMR}}
\newcommand{\repchoi}{\rep{Choi}}
\newcommand{\repmusicnn}{\rep{MusiCNN}}
\newcommand{\repjuke}{\rep{Jukebox}}

\newcommand{\etal}{et~al.}